# AQP1 Is Not Only a Water Channel: It Contributes to Cell Migration through Lin7/Beta-Catenin

Elena Monzani[1], Riccardo Bazzotti[1], Carla Perego[2], Caterina A. M. La Porta[1]*

1 Department of Biomolecular Science and Biotechnology, University of Milan, Milan, Italy, 2 Institute of General Physiology and Biological Chemistry, University of Milano, Milano, Italy

## Abstract

**Background:** AQP1 belongs to aquaporins family, water-specific, membrane-channel proteins expressed in diverse tissues. Recent papers showed that during angiogenesis, AQP1 is expressed preferentially by microvessels, favoring angiogenesis via the increase of permeability In particular, in AQP1 null mice, endothelial cell migration is impaired without altering their proliferation or adhesion. Therefore, AQP1 has been proposed as a novel promoter of tumor angiogenesis.

**Methods/Findings:** Using targeted silencing of *AQP*1 gene expression, an impairment in the organization of F-actin and a reduced migration capacity was demonstrated in human endothelial and melanoma cell lines. Interestingly, we showed, for the first time, that AQP1 co-immunoprecipitated with Lin-7. Lin7-GFP experiments confirmed co-immunoprecipitation. In addition, the knock down of AQP1 decreased the level of expression of Lin-7 and β-catenin and the inhibition of proteasome contrasted partially such a decrease.

**Conclusions/Significance:** All together, our findings show that AQP1 plays a role inside the cells through Lin-7/β-catenin interaction. Such a role of AQP1 is the same in human melanoma and endothelial cells, suggesting that AQP1 plays a global physiological role. A model is presented.

**Funding:** The authors have no support or funding to report.

**Competing Interests:** The authors have declared that no competing interests exist.

* E-mail: caterina.laporta@unimi.it

## Introduction

Aquaporins are a family of water-specific, membrane-channel proteins expressed in diverse tissues. Two functional groups of mammalian aquaporins are now recognised: aquaporins (AQP1, AQP2, AQP4, AQP5 and AQP8) which are primarily water selective and aquaglyceroporins (AQP3, AQP7, AQP9 and AQP10) which are permeable to small uncharged solutes such as lactate, glycerol and urea in addition to water [1]. The characterization of the organization of aquaporin genes and identification of their position within the human and mouse genomes have established a primary role for some aquaporins in clinical disorders such as congenital cataracts and nephrogenic diabetes insipidus. More recently, in the control of fat accumulation, aquaporins were demonstrated to play an important role [2–5]. A characterization of AQPs was carried out in neuronal stem cells [6].

Herein, we focused on AQP1 which is specifically and strongly expressed in most microvasculature endothelial cells outside the brain [7]. Recently, in AQP1 null mice, an impairment of endothelial cell migration, without altering their proliferation or adhesion, was shown [8]. In this connection, AQP1 has been proposed as a novel promoter of tumor angiogenesis [9].

Cell migration involves transient formation of membrane protrusion (lamellipodia and membrane ruffles) at the leading edge of the cell. AQP1 is localized at the leading edge and co-immunoprecipitated with several transporters involved in migration including the Na/H$^+$ and Cl$^-$/HCO3$^-$ exchangers [10]. Verkman proposed, recently, that water entry might then increase local hydrostatic pressure to cause membrane protrusion that, in turn, creates space for actin polymerization [1].

The organization of cytoskeleton requires the presence of scaffolding proteins, the plasma membrane-associated proteins containing one or several PDZ domains [11]. One scaffold complex common for epithelial and neuronal cells is the heterotrimeric complex consisting of the CASK/Lin-2, Lin-7 and Lin-10 PDZ proteins [12–15]. In mammals, Lin-7 can recruit cell adhesion molecules, receptors and signaling proteins [12–14]. CASK/Lin-2 contains bindings sites for interactions with cytoskeletal proteins [15]. On the other hand, transporters, receptors and ion channels such as the epithelial γ-aminobutyric acid (GABA) transporter (BGT-1), the N-methyl-D-aspartate (NMDA) receptor and N-type Ca$^{2+}$ channel, interact with the PDZ domains of mammalian Lin-7 [16–19]. The heterotrimeric PDZ complex plays a role in regulating the localization of interacting proteins but it is still uncertain how these components are targeted to the junction surface [17,18].

Using targeted silencing of *AQP*1 gene expression, herein, we demonstrated, for the first time, that AQP1 knock down dramatically affects the actin cytoskeleton organization through Lin-7/β-catenin interaction. As cellular model we used a melanoma cell lines which was extensively studied and character-ized by our group and that express many angiogenic/lymphoangiogenic factors (WM115) [20]. Finally, in order to assess if their effects were restricted to a specific melanoma cell line or AQP1 plays a global physiological role, we also analyzed a human endothelial cell line.

## Materials and Methods

### Cell culture

Human melanoma WM115 cell line derived from a primary epithelioid tumor (ATCC CRL 1675) was cultured in the basal medium Eagle (BME) supplemented with 10% fetal calf serum, 2 mM L-glutamine, 1% non-essential amino acids, 2% BME vitamin solution, 100 U/ml penicillin, 100 μg/ml streptomycin and 0,25 μg/ml amphotericin B (Invitrogen, Milan, Italy) as previously described [21]. Human microvascular endothelial cell line HMEC-1 was provided by Prof. E Dejana (IFOM, FIRC Molecular Oncology Institute, Milan, Italy). The HMEC-1 cell line is representative of the microvasculature and has proprieties similar to those of the original primary culture [22]. HMEC-1 cells were cultured in MCDB-131 medium (Sigma-Aldrich, Milan, Italy) supplemented with 10% FCS, 10 ng/ml epidermal growth factor (EuroClone, UK), 1 μg/ml hydrocortisone (Sigma-Aldrich, Milan, Italy) and 5% glutamine.

### siRNA synthesis

Three different AQP1 specific chemically-modified siRNA duplexes, Stealth siRNA (Invitrogen Corporation, Carlsbad, CA, USA) were designed to be homologous to the AQP1 consensus sequence (NCBI Entrez nucleotide database-accession number: M77829). Cells were transfected with HP validated negative control (CTRL1 and CTRL2), AQP1-1, AQP1-2 and AQP1-3, Sequences for siRNA were: CTRL1: sense - r(AGUCCAUACC-GUAAGCUAGUCAAUU), antisense – r(AAUUGACUAGCU-UACGGUAUGGACU); CTRL2: sense – r(UGCGUGUGUA-GAAACAGAAGGACGU), antisense – r(ACGUCCUUCUGU-UUCUACACACGCA); AQP1-1: sense (AAUGACCUGGCU-GAUGGUGUGAACU), antisense (AGUUCACACCAUCAGC-CAGGUCAUU) [23]; AQP1-2: sense(UGCUGAUGAAGACA-AAGAGGGUCGU), antisense(ACGACCCUCUUUGUCUU-CAUCAGCA); AQP1-3: sense(UUCAUCUCCACCCUGGA-GUUGAUGU), antisense(ACAUCAACUCCAGGGUGGAGA-UGAA). These siRNAs are designed with unequal stabilities of the 3′ and 5′ bases, to reduce the risk of off-target effects. Transfection for AQP1 (1, 2, 3) were carried out with a cationic lipid-based reagent i.e. lipofectamine 2000 (Invitrogen Corporation, Carlsbad, CA, USA). Tranfections were carried out according to the manufacturers' instructions. Cell plating density in a 6-well plate for tranfection with lipofectamine 2000 was $2\times10^5$/ml and a final concentration of AQP1 (1, 2 or 3) or CTRL1 and CTRL2 was 100 pmol. Transfection were carried out either in triplicates.

### GFP-Lin7

Full-length mouse Lin7A protein was fused to the C-terminus of the enhanced Green fluorescence (GFP-Lin7) by subcloning the BamH1 restriction fragment of the pGEX1-mLin7A plasmid (kindly provided by S. Kim, Standford Medical School, CA) in the mammalian expression vector pEGFP-C1 (Promega) as previously described.

### MG132 treatment

The cells were treated with MG132 (10 μmol/L, Sigma) for 1 h under the standard growth medium.

### Immunofluorescence

Cells were fixed at 4°C in 4% paraformaldehyde (methanol for co-staining with anti-AQP1 and anti- Na+/K+ ATPase antibodies) and then permeabilised with 0.01% Triton X-100 for 30 min at room temperature. Then, the cells were blocked with normal goat serum (1:20) in a solution of 1% glycine, 2% bovine serum albumine in saline for 20 min and incubated with rabbit anti-AQP1 (1:100, Alpha Diagnostic International), rabbit anti-Lin7 (1:400, Sigma), goat anti-EphB2 (1:50, Santa Cruz), mouse anti-β-catenin (1:200, BD Transduction Laboratories) or mouse anti-Na+/K+ ATPase (1:250, provided by Prof. G. Pietrini, University of Milan) antibodies overnight at 4°C. Then the cells were incubated with the respective secondary antibodies: Alexa Fluor-488-conjugated goat anti-rabbit, (1:700, Molecular Probe); TRITC-conjugated anti-goat (1:1000, Molecular Probes); Alexa Fluor-594-conjugated goat anti-mouse, (1:700, Molecular Probes) for 1h and examined on a Leika TCS NT confocal microscope. Nuclei were counterstained with 4,6-diamine-2-phenylindole dihydrochloride (DAPI; Sigma; 50 μg/ml) for 30 min.

For F-actin visualisation, the cells were fixed in 4% paraformaldehyde, permeabilised with 0.1% Triton X-100 for 30 min and incubated with 50 μg/ml FITC-conjugated phalloidin (Sigma) or Rhodamine-conjugated phalloidin (Molecular Probes) for 40 min at room temperature.

### Cell migration and invasion

$2\times10^5$ cells were plated on the upper chamber of transwell inserts and maintained in serum-free medium. The lower chamber contained medium plus 1% serum or 10% serum (chemoattractant) or no serum (negative control). Cells were incubated for 6 h at 37°C in 5% $CO_2$. After incubation, the membranes were washed briefly with saline and the non-migrated cells were scraped from the upper surface of the membrane with a cotton swab. Migrated cells remaining on the bottom surface were stained for AQP1 for wild type cells and counted 6hr later (see immunofluorescence for AQP1). For AQP1 silenced cells, the number of cells migrated were counted after haematoxylin staining. The migrated cells were counted in 20 random high-power microscope fields under ×200 magnification.

### Co-immunoprecipitation

Cells were grown to confluence, washed once with ice-cold saline and resuspended in lysis buffer (20 mM Tris-HCl pH 7.5, 2 mM EDTA, 0.5 mM EGTA, 0.5 mM PMSF, protease inhibitor cocktail, Triton 1%). Affinity purified rabbit anti-AQP1 (2 μg/ml; Alpha Diagnostic International, overnight), rabbit anti-Lin7 (1:500, Sigma, overnight), goat anti-EphB2 antibodies (2 μg/ml; Santa Cruz Biotechnology, overnight) or anti-GFP (1 μg/ml, Quantum Biotechnologies, 2hr) were added to clarified lysates (1 mg/ml). After incubation of lysates with primary antibodies, protein A-Sepharose pre-blocked with 10% bovine serum albumine was added, and the lysates were incubated for 90 min at 4°C. The immunoprecipitated samples were recovered after centrifugation (2 min at 6100 rpm at 4°C), washed three times for 2 minutes at 3100 rpm at 4°C, each with washing buffer (20 mM HEPES pH 7.5 containing 150 mM NaCl, 0.1% Triton X-100 and 10% glycerol). Immunoprecipitated proteins were eluted from Sepharose by incubation in elution buffer (62.5 mM Tris-HCl, pH 7.4, 2% w/v sodium dodecyl sulfate, 10% glycerol, β-mercaptoethanol, 0.01% bromophenol blue) and boiled for 5 min before electrophoresis. The eluted proteins were subjected to SDS/PAGE as described below. As negative control, protein A-Sepharose was pre-blocked with 10% bovine serum albumin and then incubated with the lysate (1 mg/ml) without previous incubation with the primary antibody ("pre" sample).

### Western blot analysis

Lysates (50 µg total protein per lane) were run in 10% SDS-PAGE and transferred on PVDF membrane [24]. The blots were incubated for 1–2 h in blocking solution (5% skimmed milk in Tris-buffer), then for 2h with the following primary antibodies: anti-CASK/Lin-2 (1:2000; Chemicon); anti-AQP1 (1:2000, Alpha Diagnostic International); anti-β-catenin (1:2000; Chemicon); anti-PAR-3 (1:2000; Invitrogen); anti-Lin-7 (1:2000, Sigma); anti-β-actin, (1:5000, Sigma). For anti-GFP (1:2000, Roche) the sheet was incubated overnight at 4°C. The blots were incubated for 1h with horseradish peroxidase-conjugated secondary antibodies (Amersham Pharmacia Biotech) against mouse or rabbit immunoglobulins. The bands were visualized using the ECL-Plus detection system (Amersham Pharmacia Biotech). Densitometric analysis was carried out using Image Master software (Amersham-Pharmacia Biotech, Amersham, Bucks, UK).

### RT-PCR analysis

Total cellular RNA was extracted using the RNeasy Mini Kit (Qiagen) according to the manufacturer's protocol. Complementary DNA (cDNA) was synthesized using 3–5 µg of total RNA, 2.5 µM of random nonamers, 500 µM of each of the four deoxynucleotides, and water up to 10 µl. After incubation for 10 min at 70°C and 5 minutes on ice, 10 µl of RT reaction mixture containing 1 U of enhanced avian myeloblastosis virus reverse transcriptase enzyme (eAMV-RT, Sigma), as well as 1 U of RNase inhibitor, 1× buffer for AMV-RT, and water, was added for a total volume of 20 µl. The reaction proceeded at 25°C for 15 min. and 45°C for 50 min. PCR reactions were performed in 50 µl using 200 µM dNTPs, 1.5 U of REDTaq polymerase (Sigma), 1× RED buffer, 0.3 µM of specific oligonucleotide primers, and 2 µl of cDNA. The sequence of the primers, the lengths of PCR products and the annealing temperatures were as follows: AQP1, sense 5′-TGGGGACCAAGATTTACCAA-3′ and antisense 5′-CGCAGAGTGTGGGCCACATCA, 221 bp 68°C; AQP4, sense 5′-GAATCCTCTATCTGGTCACA-3′, antisense

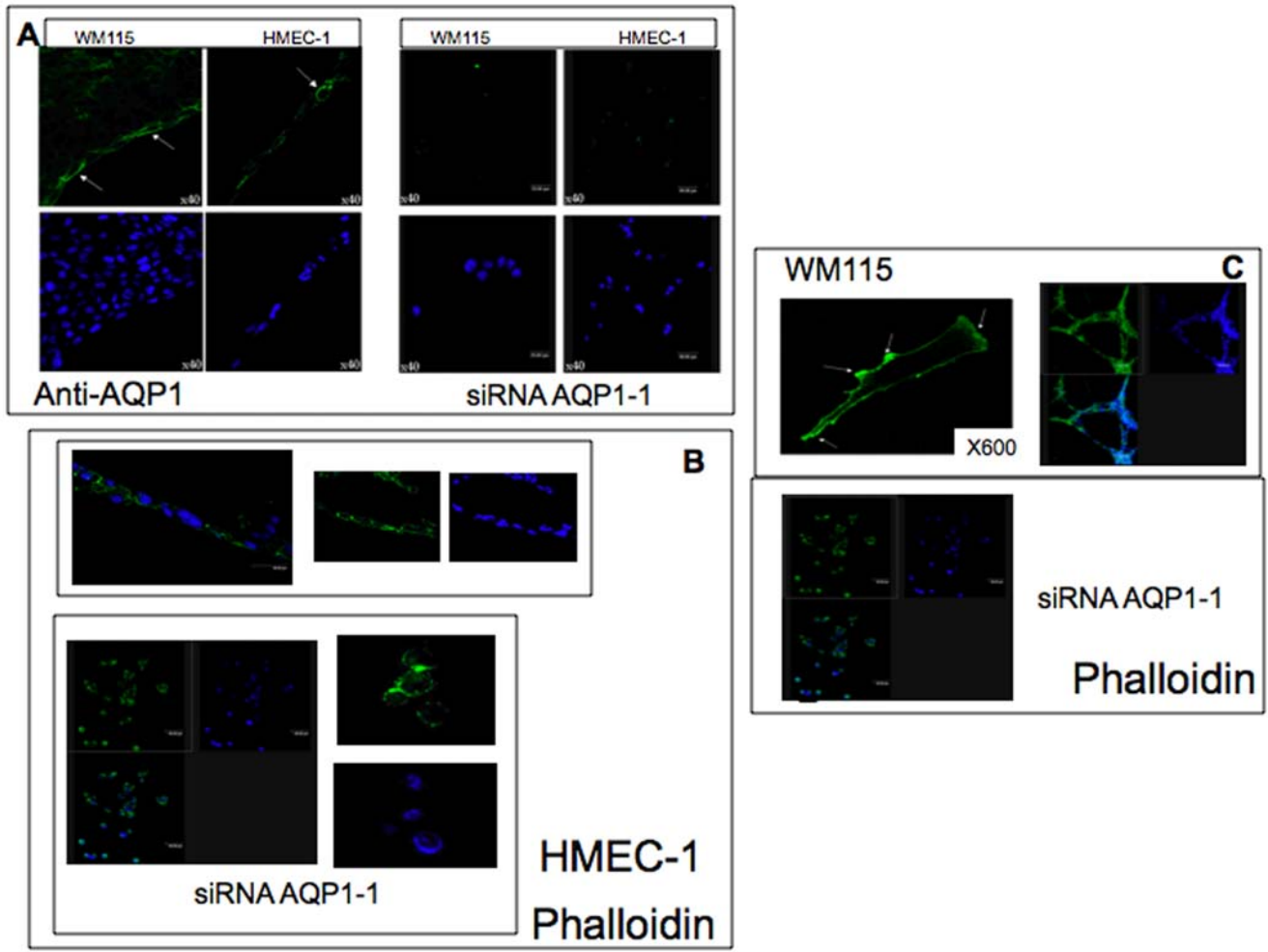

**Figure 1. Immunofluorescence of F-actin in wild type HMEC-1 () and WM115 () cells and in siRNA AQP1-1 cells.** $10\times10^5$ wild type or siRNA AQP1-1 silenced cells cells were fixed with 4% paraformaldeide and incubated with anti-AQP1 (1:100, panel A, magnification ×400) overnight and/or with phalloidin, (50ug/ml, TRITC, panel B and C) for 40 min. Nuclei are stained with DAPI (blu). HMEC-1 cells were plated on matrigel coated dishes. The figure shows that silencing of AQP1 (siRNA AQP1-1, panel A) induces a dramatically change in cell shape. (panel B and C) The latter loses the capability to organize a cord-like network.
doi:10.1371/journal.pone.0006167.g001

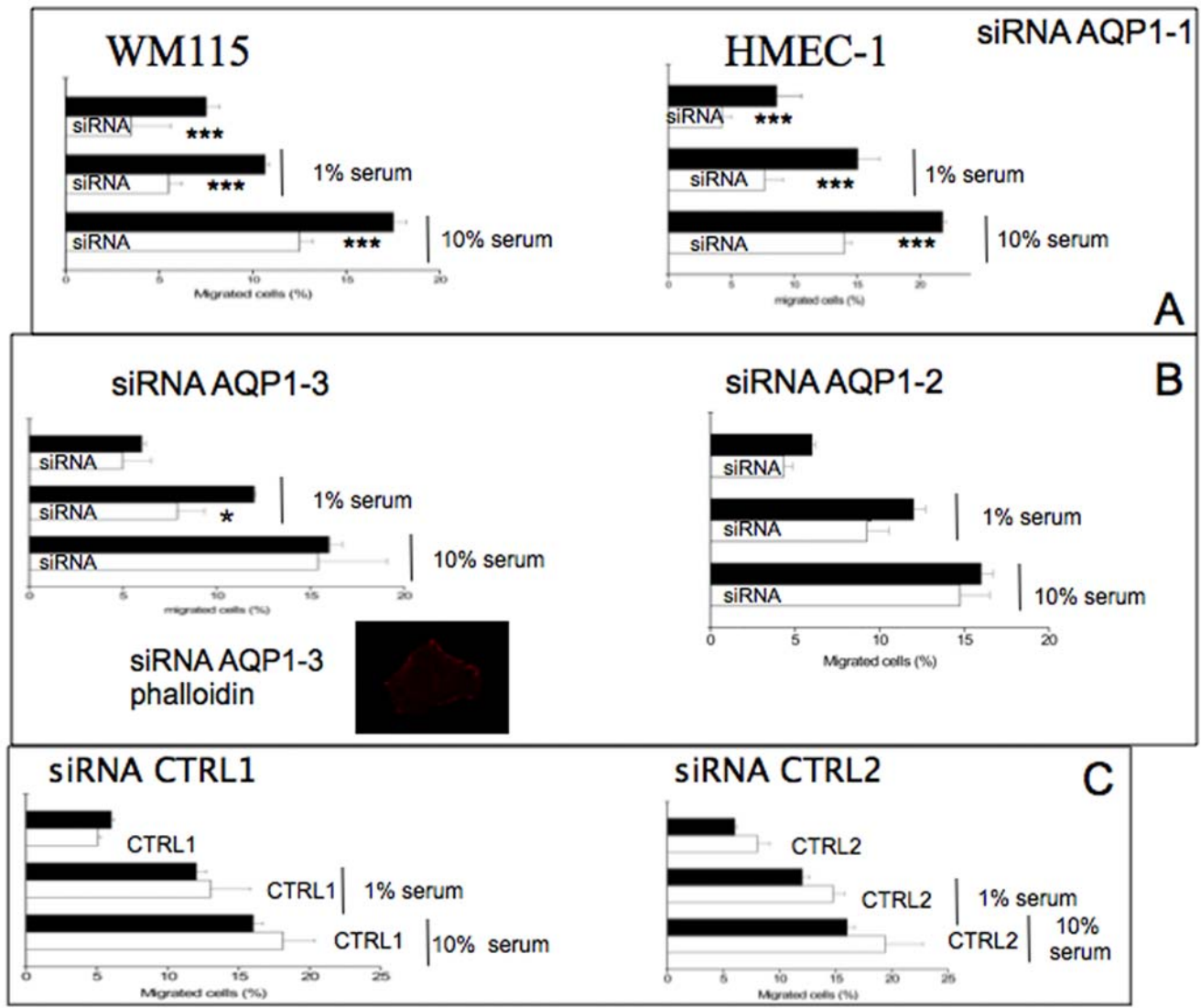

**Figure 2. Effect of siRNA AQP1-1 on cellular migration.** Panel A shows the migration of AQP1 silenced WM115 cells with siRNA AQP1-1 with respect to control cells. Parental and AQP1-1 knock-down cells were plated on a transwell polycarbonate membrane filter (6,5 mm diameter, 8 μm pores) for 6 hours. After incubation the cells were washed briefly with PBS and the non-migrated cells were scraped from the upper surface of the membrane with a cotton swab. Migrated cells remaining on the bottom surface were fixed with 4% paraformaldehyde, permeabilized with 0,01% Triton-X-100 and stained to detect AQP1. The nuclei were stained with DAPI. Percentage of transmigrated cells were obtained counting the number of cells expressing AQP1 (for parental cells) and DAPI positive cells (for AQP1 knock down cells) migrated to the underside of the filter and normalized to the total number of the cells adhering to polycarbonate membrane filter. For each experiment the magnitude of cells migration was evaluated by counting the migrated cells in 20 random high-power (×200) microscope fields. The data are the means+/−SD deviation of three independent experiment, each carried out in triplicate. **, $p<0.01$ versus control cells (black bar); *** $p<0.001$ versus control cells (black bar). Panel B shows the effect of siRNA AQP1-2 and AQP1-3 on cell migration. siRNA AQP1-3 cells were stained with phalloidin accoding to figure 1. Panel C shows the effect of siRNA CTRL1 and CTRL2 on the migration capacity of WM115 cells according to methods described in panel A. * $p<0.05$ versus wild type cells. The figure shows that siRNA AQP1-1 induced a significant reduction in the migratory capability in both cell lines. In contrast, siRNA AQP1-3 does not induce significant effects on the migration as well as on the shape of the cells. To confirm the specificity of the results, we transfected the cells with CTRL 1 and CTRL2, both do not modify significantly the migration capacity.
doi:10.1371/journal.pone.0006167.g002

AQP4 5′-TGTTTGCTGGGCAGCTTTGCT-3′ 429 bp 92°C; AQP9, sense 5′-GCCTAGAGCCCATTGCCATCG-3′, reverse AQP9 5′-TGGTTTGTCCTCAGATTGTTC-3′, 289 bp 94°C; Lin-2 sense 5′-CTAGCCGCTGTGTCAAGTCA-3′, antisense 5′-TTCTTCCAATACCTCTTTGGC 321 bp 60°C; GAPDH, sense 5′- TCTAGACGGCAGGTCAGGTCCACC, reverse 5′-CCACCCATGGCAAATTCCATGGCA 589 bp, 58°C. The PCR products were resolved by electrophoresis in 1.5% agarose gels containing ethidium bromide and photographed under UV light.

## Results

### Knock down of AQP 1 in HMEC-1 and WM115 cell lines induces changes in cell shape and F-actin organization

WM115 and HMEC-1 cells express mainly AQP1 (see Supplementary materials, Fig. S1). We knocked down the AQP1 water channel using three different siRNA: AQP1-1, AQP1-2, AQP1-3 in both WM115 and HMEC-1 cells (see Supplementary materials, Fig. S2). A significant suppression of AQP1 (90%) occurred in WM115 tranfected cells with siRNA AQP1-1 and with siRNA AQP1-2 (72%) but not with AQP1-3 (23%) (Supplementary materials, Fig. S2). Furthermore, siRNA CTRL1 and 2 did not modify the level of expression of AQP1 (Supplementary Materials, Fig. S2). Interestingly, in both cell lines, the silencing of AQP1 (siRNA AQP1-1) induced a dramatically change in cell shape. In non-silenced cells, F-actin was preferentially polarized at the leading edge of the plasma membrane and the cells were able to organize a cord-like network in vitro. In contrast, both AQP1-1 silenced cells, were rounded up (Fig. 1, panel B and C). The integrity of the F-actin cytoskeleton is critically dependent on cell volume changes and several volume-regulated transport proteins including water channels exhibit functional interaction with the cytoskeleton playing a pivotal role in cell volume regulation [1]. The ability to establish and maintain a polarized state of the cell, that is the basis for migration, was suggested to be reflected by a polarized distribution of the ion and water channels [24,25] that determine the water influx, providing the force that extend the plasma membrane [26] and that promotes the actin polymerization in the front part of cells formed by lamellipodium [27]. Our data shows an AQP1 dependent organization of cell shape (Fig. 1).

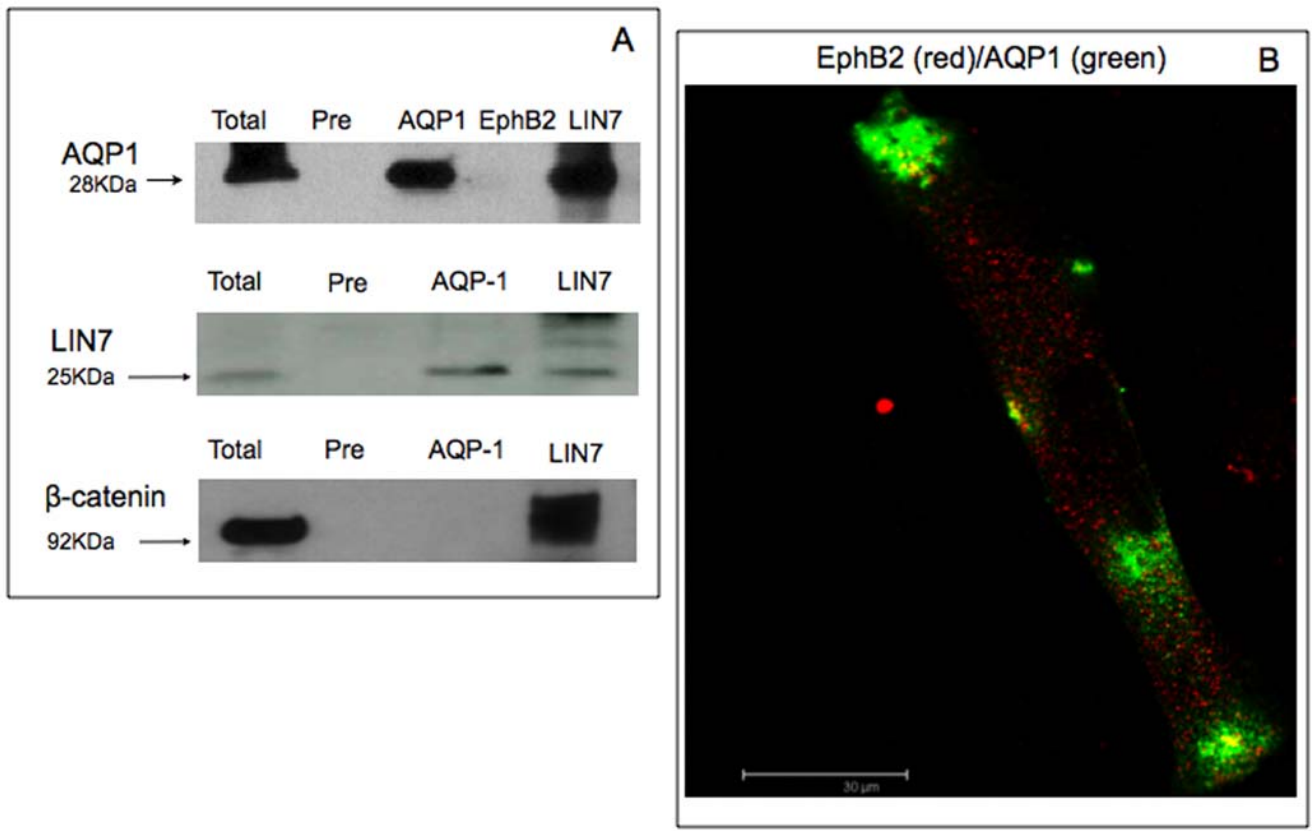

**Figure 3. AQP1 co-immuoprecipitation with Lin7.** Panel A. Co-immunoprecipitation analysis. 1 mg/ml of total extracts were immunoprecipitated with 2 μg/ml of anti-AQP-1 or 1:500 anti-Lin-7 or 2 μg/ml anti-EphB2. The samples were then loaded onto 10% SDS-PAGE gel and immunoblotted on PVDF sheet. The latter was incubated with anti- AQP-1 (1:2000) or anti-Lin7 (1:2000) for 2h. Protein A-Sepharose pre-blocked with 10% of BSA and then incubated with the lysate without previous incubation with the primary antibody was used as a negative control (Pre). Panel B. Double immunofluorescence of EphB2 and AQP1. $2\times10^5$ cells were plated into glass coverslip and fixed at 4°C for 20 min in 4% paraformaldehyde. Then, the cells were permeabilized with 0,01% Triton-X-100, blocked for 30 min with normal goat serum (1:20) and incubated with primary antibody anti-EphB2 (1:50) and anti-AQP1 (1:100) overnight. The cells were finally incubated with secondary antibodies conjugated with Alexa 594-conjugated goat anti-mouse (red) or Alexa Fluor-488-conjugated goat anti-rabbit (green) (1:700) and the nuclei stained with DAPI. The figure shows that AQP1 co-immunprecipitates with Lin-7 and beta-catenin but not with EphrB2. Accordingly, the latter does not co-localized with AQP1.
doi:10.1371/journal.pone.0006167.g003

**AQP1 affects migration potential of WM115 and HMEC-1 cells.** Since AQP1 was demonstrated to be involved in angiogenesis [8], we investigated the effect of this water channel on cell migration in WM115 and HMEC-1 cells. As shown in Figure 2A, AQP1 knock down (siRNA AQP1-1) induced a significant reduction in the migratory capability of both cell lines. In contrast, siRNA AQP1-3 did not induce any significant effects on the migration or on the shape of the cells. Accordingly, siRNA AQP1-2 (which gave an intermediate effect on AQP1 expression) showed a little effect on cell migration (Fig. 2B). On the other hand, CTRL 1 and 2 did not modify significantly the migration capacity of both cell lines (Fig. 2C)

**Lin-7 co-immunoprecipitates with AQP1 in HMEC-1 and WM115 cells.** Lin-7 is a component of the cadherin-catenin complex through the physical interaction with the PDZ target sequence of beta-catenin [28]. Recently, some transportes, receptors and ion channels are demonstrated to interact with the PDZ domains of mammalian Lin-7 [16-18]. To verify whether Lin-7 is a component of AQP1-containing complex, we tested the association of these proteins by co-immunoprecipitation (Fig. 3). Furthermore, the polyclonal Lin-7 antibody (but not a pre-immune serum) co-immunoprecipitated

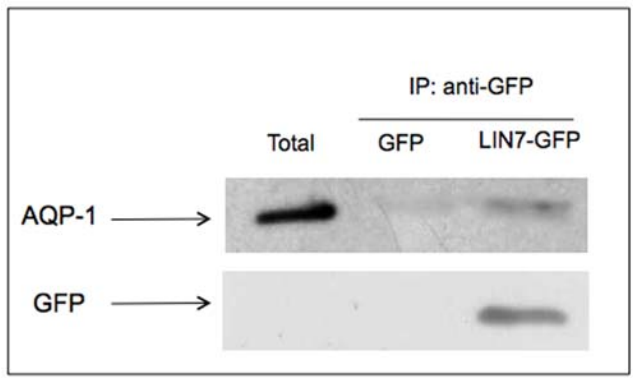

**Figure 4. GFP-Lin7 co-immunoprecipitate with AQP1.** 1 mg/ml of total extract of the WM115 cells expressing the LIN7-GFP fusion protein or GFP was incubated for 2h at 4°C with the anti-GFP (1microg/µl). Protein A-Sepharose pre-blocked with 10% bovine serum albumine was added and the samples were incubated for 90 min at 4°C. The immunoprecipitated samples recovered by centrifugation, were loaded onto 10% SDS-PAGE gel, immunoblotted on PVDF sheet and incubated with anti-AQP-1 (1:2000) for 2h or anti-GFP (1:2000) overnight. The figure shows a typical experiment out of the three independent experiments carried out (all giving the same results). The figure shows that in cells expressing LIN7-GFP fusion protein, LIN7-GFP co-localized with AQP1. In both total and GFP transfected cells, no AQP1 band is detected.
doi:10.1371/journal.pone.0006167.g004

AQP1 from both WM115 and HMEC-1 lysates (Fig. 3A). We have also verified that Lin-7 and beta-catenin co-immunoprecipitated in our cells according to previous paper [28] (Fig. 3A). In contrast, Ephrin B2, a protein involved in the maintaining of morphology of the cells and in various biological process such as angiogenesis, did not co-immunoprecipate or co-immunolocalize with AQP1 (Fig. 3A and B). Since Lin-7 interacting with Lin-2/CASK bind cytoskeletal proteins [15], we investigated whether the two cell line expressed Lin-2/CASK. However, we found no significant expression of Lin-2/CASK by RT-PCR or immunoblot analyses (not shown).

Finally, in figure 4 is shown that AQP1 is retained specifically by the Lin7-GFP, confirming co-immunoprecipitation experiments.

**Knock down of AQP1 affects Lin-7/β-catenin expression.** Since Lin-7 co-immunoprecipitates with AQP1, we knocked down AQP1 with three different siRNA (AQP1-1, AQP1-2 and AQP1-3) and two CTRLs and investigated what happens to Lin-7, beta-catenin and beta-actin (housekeeping gene) in both cell lines (Fig. 5). Furthermore, we have also analysed the effect of silencing AQP1 on PAR-3, another PDZ-containing protein critical for cell polarity localized at the intracellular junctions of endothelial cells [29]. Negative control (CTRL1 and 2) did not affect significantly, Lin7, beta-catenin or AQP1 (Fig. 5A). Accordingly, siRNA AQP1-3 (which is not able to reduce significantly the level of expression of AQP1 and did not modify the morphology of the cells, Fig. 2B) did not modify beta-catenin or Lin-7 expression (Fig. 5B). siRNA AQP1-2 gave an intermediate effect (Fig. 5B).

Importantly, the treatment of AQP1 silenced cells with the proteasome inhibitor MG132 enhanced the recovery of beta-catenin (70%, Fig. 6A) and partially of Lin-7 (16%, data not shown). Accordingly, in cells transfected with CTRL1 no effect after MG132 treatment occurred (Fig. 6B).

## Discussion

Aquaporins (AQPs) are the integral plasma membrane proteins involved in water transport in many fluid-transporting tissues. Interestingly, in AQP1 null mice endothelial cell migration was impaired, and *in vivo* angiogenesis reduced [8].

Herein, we show, for the first time, in two different cell line expressing AQP1, a human melanoma cell line WM115 and a human microvascular cell line HMEC-1 that the knock down of AQP1 induced a re-organization of F-actin and affected the cell shape. In non-silenced cells F-actin was preferentially polarized at the leading edge of the plasma membrane and the cells were able to organize a pseudovascular network. In contrast, siRNA AQP1-1 cells have lost the capability to organize such a network. From the functional point of view, AQP1 silenced cells decreased their capability to migrate. In this connection, a remarkably impaired growth of implanted tumors was demonstrated in AQP1 null mice, with reduced tumor vascularity and an extensive necrosis [8]. Interestingly, in AQP-1 null mice cell migration of melanoma cells was greatly impaired with abnormal *in vitro* vessel formation [8].

The novelties of the present paper are the following. Firstly, AQP1 plays the same role in human melanoma and endothelial cells, suggesting that this water channel has a global physiological role. Secondly, AQP1 interacts at least, with Lin-7/beta catenin signaling. All these aspects are particular intriguing. Actually, the polarization at the leading edge of migrating cells has been demonstrated for several transporters involved in migration, including $Na^+/H^+$ and $Cl^-/HCO_3^-$ exchangers and the $Na^+/HCO_3^-$ co-transporter [30], and cell migration involves the transient formation of membrane protrusions (lamellipodia and plasma membrane ruffles) at the leading edge of the cell, suggesting that a rapid local changes in ion fluxes and cell volume are accompanied by rapid transmembrane water movement [8]. Thereby, the actin cleavage and ion uptake at the tip of lamellipodium might create local osmotic gradients that drive the influx of water across the plasma membrane [10,31–33]. Saadoun et al. postulated that water entry increases local hydrostatic pressure causing the polarization of AQP1 [8]. However, at yet, the intracellular mechanisms triggered by AQP1 are not well known. Herein we show, for the first time, that Lin-7 co-immunoprecipitates with AQP1 and interacts with beta-catenin through Lin-7, thereby affecting the organization of cytoskeleton. Intriguingly, our experiments with the proteosomal inhibitor imply that lack of AQP1 targets the Lin-7/beta-catenin complex to proteolytic degradation.

All together, these data expand the physiological role of AQP1 beyond its water transport function. This protein rather emerges as a critical scaffold for a plasma membrane associated multi-protein complex important for cytoskeleton build-up, adhesion and motility. A model for these functions of AQP1 is presented in Figure 7. In this model we focused the role of AQP1 in the stabilization of cytoskeleton complex through the involvement, at least, of Lin-7/beta-catenin.

All of all, our findings corroborate the analysis of manifold cellular functions of AQPs in normal cells and in diseases and the possibility to consider aquaporins as specific therapeutic targets for various pathophysiological conditions [34].

## Supporting Information

**Figure S1** Expression of AQPs and intracellular localization. Panel A. RT-PCR of AQP1 in WM115 and HMEC-1 cells. Total RNA was purified from subconfluent cells using the RNeasy Mini Kit from Qiagen. One microgram of total RNA was reverse

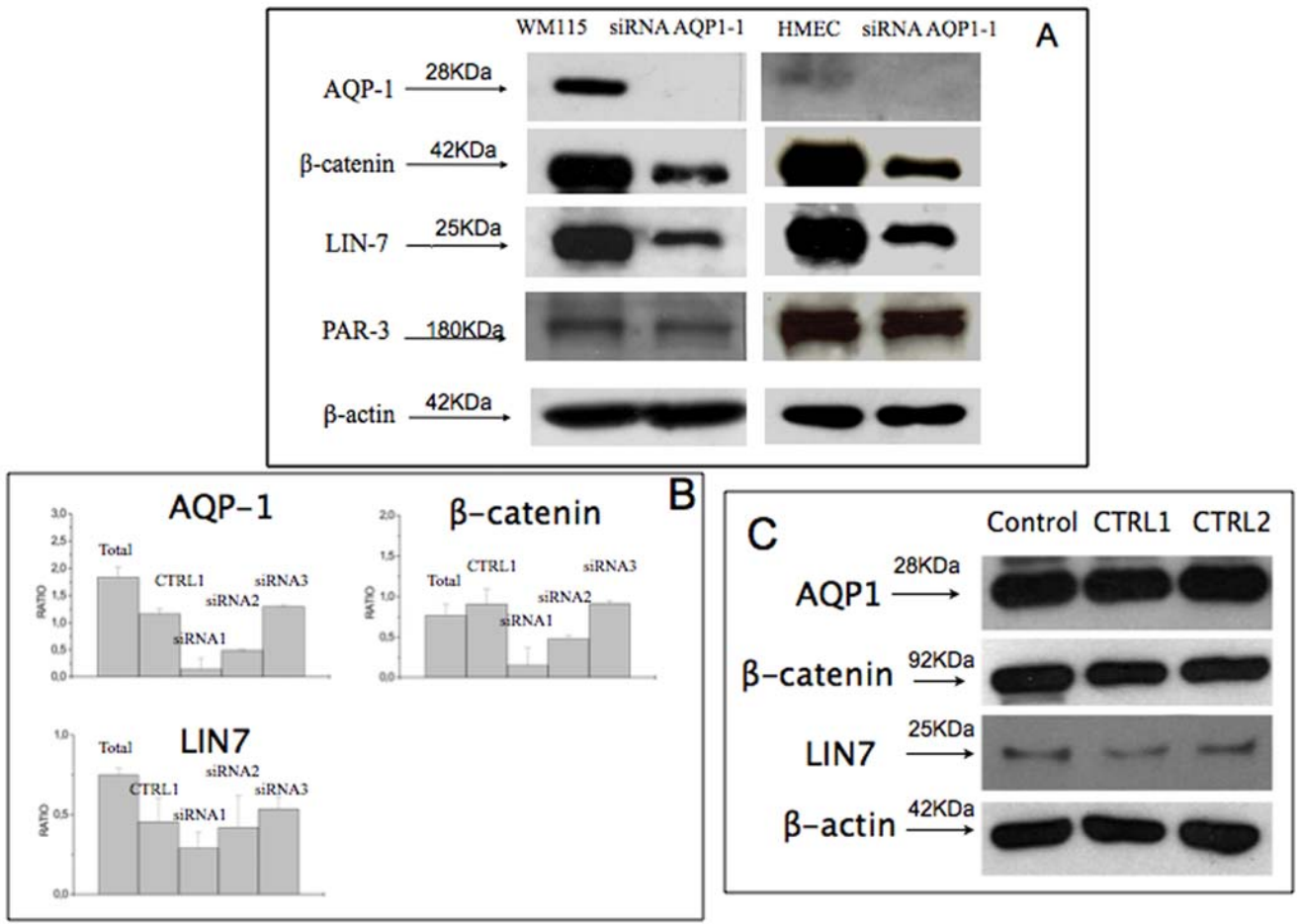

**Figure 5. Effect of siRNA AQP1 on Lin7, beta-catenin expression.** Panel A. Western blot analysis of AQP1, beta-catenin, LIN-7 and beta-actin in wild type WM115 or HMec-1 cells and transfected with siRNA AQP1-1. 50microg protein were loaded on 10% SDS PAGE and transferred on PVDF sheet. The latter was incubated for 2h with anti-AQP-1 (1:2000); anti-PAR-3 (1:2000), anti-beta-catenin (1:2000), anti-Lin-7 (1:2000); anti-beta-actin, (1:5000). The bands were visualized using the ECL-Plus detection system. Panel B. Ratio of densitometric values for each line with respect to beta-actin of AQP1, beta-catenin, Lin-7 in WM115 cells untreated or transfected with siRNA AQP1-1 or AQP1-2 or AQP1-3 according to Materials and Methods section (details in panel A). Panel C. Western blot of AQP1, beta-catenin, LIN-7 and beta-actin in WM115 cells transfected with siRNA CTRL1 or CTRL2 according to Materials and Methods section (details in panel A). The figure shows a decreased expression of Lin7 and beta-catenin in siRNA AQP1-1 transfected cells. No changes occur in CTRL1 or CTRL2 transfected cells. Moreover, no change of PAR3 expression in AQP1-1 silenced cells occurs.
doi:10.1371/journal.pone.0006167.g005

transcribed and amplified by Enhanced avian hs RT-PCR (Sigma) according to the manufacturer's instructions. PCR conditions and primers are reported in the Materials and Methods section. As housekeeping gene, GAPDH was used. In order to check the specificity of the amplified bands they are sequenced. Panel B. RT-PCR of AQP 8, 4 and 9 in WM115 cells for HMEC-1 and WM115 cells as described in panel A. As positive control, human brain tissue expressing AQP4 and AQP8 was used. Panel C. Co-immunofluorescence of AQP1 and Na+/K+ ATPase in WM115 cells. The cells fixed with methanol for 10 min and then permeabilized with 0.01% TRITON X-100 for 20 min, incubated with the primary antibodies (anti-AQP-1, 1:100; mouse anti-Na+/K+ ATPase (1:250) for 2h at room temperature. The cells were then incubated with secondary antibodies: Alexa Fluor-488 conjugated goat anti-rabbit, (1:700, Molecular Probe), Alexa Fluor-594-conjugated goat anti-mouse, (1:700, Molecular Probes) for 1h and examined on a Leika TCS NT confocal microscope. The localization of both proteins in plasma membrane are shown x-y and z-stack analysis. AQP1 is the main aquaporins expressed in these cells. The protein is mainly expressed at the level of plasma membrane. In order to clearly demonstrate the plasma membrane localization of AQP1, we have co-stained the cells with anti-AQP1 and anti-Na+/K+ ATPase (typical membrane marker) antibodies. As shown in panel C of figure 1, AQP1 shows a plasma membrane localization as demonstrated both by the merge and Z-stack analysis.
Found at: doi:10.1371/journal.pone.0006167.s001 (0.14 MB TIF)

**Figure S2** AQP1 siRNAs in WM115 and HMEC-1 cells. WM115 or HMec-1 cells were tranfected with siRNA AQP1-1, AQP1-2, AQP1-3 or negative control siRNA CTRL1 for 48 h as described in the Materials and Methods section. 50ug protein was submitted to 10% SDS-PAGE and transferred to a PVDF sheet. The latter was incubated with anti-AQP1 (1:2000) or beta-actin (1:5000) as housekeeping protein, for 2h. Then the sheet was incubated with secondary antibody and visualized using the ECL detection system. Densitometric analysis expressed as ratio of each line with respect to beta-actin is shown. The figure shows that

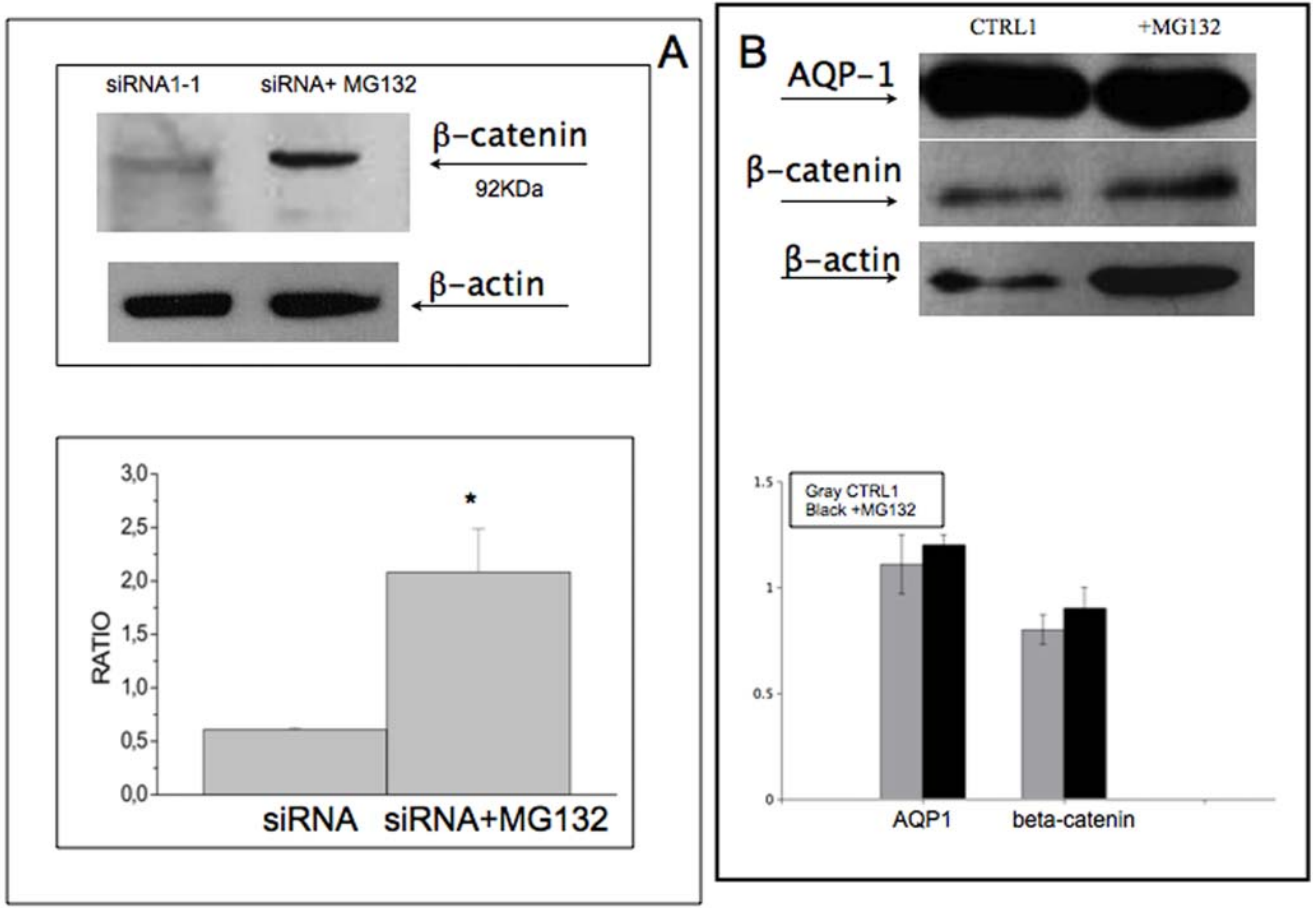

**Figure 6. Effect of MG132 on the expression of beta-catenin in siRNA AQP1 cells.** Panel A. Analysis of proteasome activity in WM115 cells. siRNA AQP1-1 tranfected cells were incubated with MG132 (10uM) for 1h and then the levels of expression of beta-catenin and beta-actin were measured by western blot as described in figure 7. A typical experiment and the ratio of densitometric value for each line with respect to beta-actin of three independent experiments are shown. Panel B shows a typical western blot and densitometric analysis of cells transfected with siRNA CTRL-1 and after the treatment with MG132 (10uM) on AQP1, beta-catenin expression. beta actin was used as housekeeping protein. The figure shows that the treatment with MG132 recoveries beta-catenin expression.
doi:10.1371/journal.pone.0006167.g006

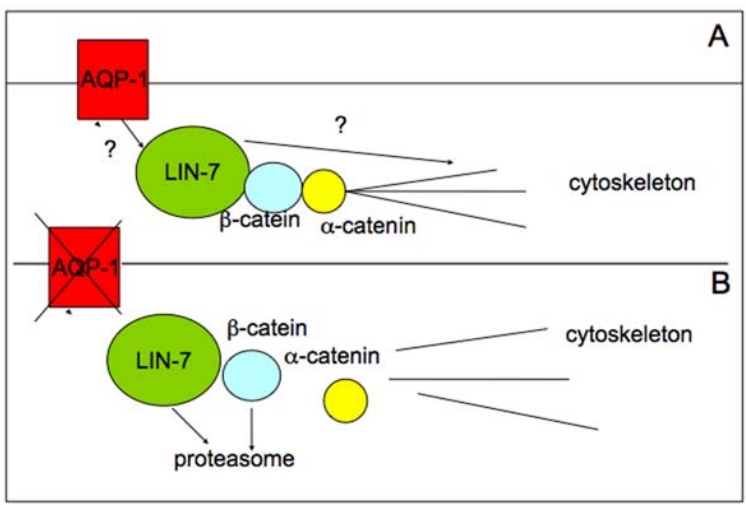

**Figure 7. Model of interaction of AQP1 with the cytoskeleton.** Proposed model. Panel A. The interaction between AQP1 and Lin7 probably mediated by other factors induces a local organization of an adhesion system that mediates the organization of F-actin. Panel B. The missing of AQP1 leads to a destabilization of the complex Lin-7/beta-catenin/F-actin as well as of other factors leading to a re-organization of F-actin and from the functional point of view leads to a reduced migration and invasive capacity.
doi:10.1371/journal.pone.0006167.g007

siRNA1-1 is more effective to down-regulate the expression of AQP1 with respect to AQP1-2 and AQP1-3.
Found at: doi:10.1371/journal.pone.0006167.s002 (0.12 MB TIF)

## Author Contributions

Conceived and designed the experiments: CAMLP. Performed the experiments: EM RB. Analyzed the data: EM RB CAMLP. Contributed reagents/materials/analysis tools: CP. Wrote the paper: EM RB CAMLP.

## References

1. Verkman AS (2007) More than just water channels: unexpected cellular roles of aquaporins. J Cell Sci 118: 3225–3232.
2. King LS, Kozovo D, Agre P (2002) From structure to disease: the evolving tale of aquaporin biology. Nat Rev Mol Cell Biol 5: 687–698.
3. Oshio K, Binder DK, Liang Y, Bollen A, Feuerstein B, et al. (2005) Expression of the Aquaporin-1 Water Channel in Human Glial Tumors. Neurosurgery 56: 375–381.
4. Rodríguez A, Catalán V, Gómez-Ambrosi J, Frühbeck G (2006) Role of aquaporin-7 in the pathophysiological control of fat accumulation in mice. FEBS Lett 580: 4771–4776.
5. Agre P (2006) The aquaporin water channels. Proc Am Thorac Soc 3: 5–13.
6. Cavazzin C, Ferrari D, Facchetti F, Russignan A, Vescovi AL, et al. (2007) Unique expression localization of aquaporin- 4 and aquaporin-9 in murine and human neural stem cells and in their glial progeny. Glia 53: 167–181.
7. Endo M, Jain RK, Witwer B, Brown D (1999) Water channel (aquaporin 1) expression and distribution in mammary carcinomas and glioblastomas. Microvasc Res 58: 89–98.
8. Saadoun S, Papadopoulos MC, Hara-Chikuma M, Verkman AS (2005) Impairment of angiogenesis and cell migration by targeted aquaporin-1 gene disruption. Nature 434: 786–792.
9. Clapp C, Martínez de la Escalera G (2006) Aquaporin-1: a novel promoter of tumor angiogenesis. Trends Endocrinol Metab 17: 1–2.
10. Schwab A (2001) Function and spatial distribution of ion channels and transporters in cell migration. Am J Physiol Renal Physiol 280: F739–F747.
11. Craven SE, Bredt DS (1998) PDZ proteins organize synaptic signaling pathways. Cell 93: 495–498.
12. Borg JP, Straight SW, Kaech SM, de Taddéo-Borg M, Kroon DE, et al. (1998) Identification of an evolutionarily conserved heterotrimeric protein complex involved in protein targeting. J Biol Chem 273: 31633–31636.
13. Butz S, Okamoto M, Südhof TC (1998) A tripartite protein complex with the potential to couple synaptic vesicle exocytosis to cell adhesion in brain. Cell 94: 773–782.
14. Kaech SM, Whitfield CW, Kim SK (1998) The LIN-2/LIN-7/LIN-10 complex mediates basolateral membrane localization of the C. elegans EGF receptor LET-23 in vulval epithelial cells. Cell 94: 761–771.
15. Cohen AR, Woods DF, Marfatia SM, Walther Z, Chishti AH, et al. (1998) Human CASK/LIN-2 binds syndecan-2 and protein 4.1 and localises to the basolateral membrane of epithelial cells. J Cell Biol 142: 129–138.
16. Jo K, Derin R, Li M, Bredt DS (1999) Characterization of MALS/Velis-1, -2, and -3: a family of mammalian LIN-7 homologs enriched at brain synapses in association with the postsynaptic density-95/NMDA receptor postsynaptic complex. J Neurosci 19: 4189–4199.
17. Maximov A, Südhof TC, Bezprozvanny I (1999) Association of neuronal calcium channels with modular adaptor proteins. J Biol Chem 274: 24453–24456.
18. Perego C, Vanoni C, Villa A, Longhi R, Kaech SM, et al. (1999) PDZ-mediated interactions retain the epithelial GABA transporter on the basolateral surface of polarized epithelial cells. EMBO J 18: 2384–2393.
19. Simske JS, Kaech SM, Harp SA, Kim SK (1996) LET-23 receptor localization by the cell junction protein LIN-7 during C. elegans vulval induction. Cell 85: 195–204.
20. ElenaMonzani, FlorianaFacchetti, EnricoGalmozzi, ElenaCorsini, AnnaBenetti, et al. (2007) Melanoma contains CD133 and ABCG2 positive cells with enhanced tumorigenic potential. Eur J Cancer 43: 935–946.
21. La Porta CA, Porro D, Comolli R (2002) Higher levels of melanin and inhibition of cdk2 activity in primary human melanoma cells WM115 overexpressing nPKCdelta. Melanoma Res 12: 297–307.
22. Ades EW, Candal FJ, Swerlick RA, George VG, Summers S, et al. (1992) HMEC-1: establishment of an immortalized human microvascular endothelial cell line. J Invest Dermatol 99: 683–690.
23. Towbin H, Staehelin T, Gordon J (1979) Electrophoretic transfer of proteins from polyacrylamide gels to nitrocellulose sheets: procedure and some applications. PNAS 76: 4350–4354.
24. Monzani E, Shtil A, La Porta CAM (2008) The water channels, new druggable targets to combat cancer cell survival, invasiveness and metastasis. Curr Drug Targets 8: 1132–1137.
25. Papdopoulos MC, Saadoun S, Verkman AS (2008) Aquporins and cell migration. Pfuger Arch 456: 693–700.
26. Schwab A (2001) Function and spatial distribution of ion channels and transporters in cell migration. Am J Physiol Renal Physiol 280: F739–F747.
27. Schwab A, Nechyporik-Zloy V, Fabian A (2007) Stock cells move when ions and water flow. Pfuger Arch 453: 421–432.
28. Perego C, Vanoni C, Massari S, Longhi R, Pietrini G (2000) Mammalian LIN-7 PDZ proteins associate with beta-catenin at the cell-cell junctions of epithelia and neurons. EMBO J 19: 3978–3989.
29. Ebnet K, Aurrand-Lions M, Kuhn A, Kiefer F, Butz S, et al. (2003) The junctional adhesion molecule (JAM) family members JAM-2 and JAM-3 associate with the cell polarity protein PAR-3: a possible role for JAMs in endothelial cell polarity. J Cell Sci 116: 3879–3891.
30. Schwab A, Schuricht B, Seeger P, Reinhardt J, Dartsch PC (1999) Migration of transformed renal epithelial cells is regulated by K+ channel modulation of actin cytoskeleton and cell volume. Pflugers Arch 438: 330–337.
31. Rosengren S, Henson PM, Worthen GS (1994) Migration-associated volume changes in neutrophils facilitate the migratory process in vitro. Am J Physiol 267: C1623–C1632.
32. Lauffenburger DA, Horwitz AF (1996) Cell migration: a physically integrated molecular process. Cell 84: 359–369.
33. Condeelis J (1993) Life at the leading edge: the formation of cell protrusions. Ann Rev Cell Biol 9: 411–444.
34. Landon S, Kozono D, Agre P (2004) From structuring to disease: the evolving tale of aquaporin biology. Nature Reviews Molecular Biology 5: 687–698.